\documentclass[aps,preprint,showpacs,superscriptaddress]{revtex4-1}  
\usepackage{graphicx}  
\usepackage{dcolumn}   
\usepackage{bm}        
\usepackage{amssymb}   
\usepackage[]{natbib}
\usepackage{hyperref}
 \usepackage[top=.75in, bottom=1in, left=1in, right=1in]{geometry} 

\usepackage{color}

\newcommand{\reef}[1]{(\ref{#1})}
\newcommand{\be}{\begin{eqnarray}}
\newcommand{\ee}{\end{eqnarray}}
\newcommand{\bea}{\begin{eqnarray}}
\newcommand{\eea}{\end{eqnarray}}

\newcommand{\vareps}{\varepsilon}
\newcommand{\eps}{\epsilon}
\def\beq{\begin{equation}} 
\def\eeq{\end{equation}} 
\def\<{\langle}
\def\>{\rangle}

\newcounter{page0}
\renewcommand{\thepage}
{\setcounter{page0}{\value{page}}
\addtocounter{page0}{-1}
\ifnum \value{page0}<10
0.\arabic{page0}
\else
\addtocounter{page0}{-10}
1.\arabic{page0}
\fi
}

\begin{document}

 
\hfill CERN-PH-TH/2013-219

\title{Conformal Field Theories in Fractional Dimensions}

\author{Sheer El-Showk}
\affiliation{Institut de Physique Th\'eorique CEA Saclay, CNRS-URA 2306, 91191 Gif sur Yvette, France}
\address{Centre de Physique Th\'eorique, \'Ecole Polytechnique, CNRS, Palaiseau, France}
\affiliation{CERN, Theory Division, Geneva, Switzerland}

\author{Miguel Paulos}    
\affiliation{Department of Physics, Brown University, Box 1843, Providence, RI 02912-1843, USA}

\author{David Poland}    
\affiliation{Department of Physics, Yale University, New Haven, CT 06520, USA}

\author{\\Slava Rychkov}    
\affiliation{CERN, Theory Division, Geneva, Switzerland}
\affiliation{Laboratoire de Physique Th\'eorique, \'Ecole Normale Sup\'erieure\\ \& Facult\'e de Physique, Universit\'e Pierre et Marie Curie, Paris, France} 

\author{David Simmons-Duffin}    
\affiliation{School of Natural Sciences, Institute for Advanced Study, Princeon, New Jersey 08540, USA} 

\author{Alessandro Vichi}    
\affiliation{Theoretical Physics Group, Ernest Orlando Lawrence Berkeley National Laboratory\\
 \& Center for Theoretical Physics, University of California, Berkeley, CA 94720, USA \vspace{.5cm}}   


\begin{abstract}
We study the conformal bootstrap in fractional space-time dimensions, obtaining rigorous bounds on operator dimensions. 
Our results show strong evidence that there is a family of unitary Conformal Field Theories connecting the 2D Ising model, the 3D Ising model, and the free scalar theory in 4D. We give numerical predictions for the leading operator dimensions and central charge in this family at different values of $D$ and compare these to calculations of $\phi^4$ theory in the $\vareps$-expansion.
\end{abstract}

\maketitle

\section{\label{sec:level1} Introduction }

The past few years have seen great progress in our understanding of Conformal Field Theories (CFTs), particularly in 3 and 4 dimensions. The numerical analyses performed in 
\cite{Rattazzi:2008pe,Rychkov:2009ij,Caracciolo:2009bx,Poland:2010wg,Rattazzi:2010gj,Rattazzi:2010yc,Vichi:2011ux,Poland:2011ey,Rychkov:2011et,Liendo:2012hy,ElShowk:2012ht,ElShowk:2012hu,Beem:2013qxa,Kos:2013tga,Gliozzi:2013ysa} have clearly demonstrated that the `conformal bootstrap' constraints of unitarity and crossing symmetry \cite{Ferrara:1973yt,Polyakov:1974gs} impose severe constraints on CFTs. Moreover, certain special \mbox{theories} (such as the 3D Ising model) appear to saturate these constraints.  On the other hand, following recent advances in our understanding of conformal blocks \cite{DO1,DO2,DO3,ElShowk:2012ht,Hogervorst:2013sma,Hogervorst:2013kva,Fitzpatrick:2013sya,Kos:2013tga}, as well as analytic studies of the bootstrap \cite{Heemskerk:2009pn,Fitzpatrick:2012yx,Komargodski:2012ek}, it has become transparent that the space-time dimension is simply a parameter which enters the bootstrap constraints. An analytic continuation to non-integer space-time dimension can be done in a completely straightforward way.

It is then natural to ask, does crossing symmetry have anything to say about the space of CFTs in non-integer dimensions? Can we find a family of solutions to the crossing symmetry constraint that interpolates between the 2D Ising model, the 3D Ising model, and the 4D free scalar? The purpose of this paper is to start addressing these questions.

\section{\label{sec:level2}CFT in fractional dimensions}

The notion of non-integer dimensions is not new to quantum field theory. A widely used method to regularize the perturbative expansion of quantum field theories is dimensional regularization---analytically continuing Feynman integrals to non-integer dimensions. In this case the analytic continuation is just a computational trick. Wilson and Fisher \cite{Wilson:1971dc,Wilson:1972cf} were the first to use such a continuation to connect theories living in different integer dimensions. They focused on $\phi^4$ theory in $D<4$ dimensions, which for $\vareps=4-D\ll 1$ has a weakly-coupled infrared fixed point. Analytically continuing this family of fixed points to $\vareps=1$ and $2$ should give, they argued, the infrared fixed point of the 3D and 2D Ising models. This observation is by now widely accepted and became the basis of the $\vareps$-expansion technique for computing the critical exponents of strongly coupled models. The results of the $\vareps$-expansion agree well with other approximation schemes, Monte-Carlo simulations, and exact results when available. This strongly suggests that the basic idea is correct, in spite of the fact that it has never been justified beyond perturbation theory, and even a proper definition of what it means to have a field theory in non-integer $D$ has not been given.

In this paper we will provide new evidence for the existence of a line of fixed points interpolating between 2 and 4 dimensions which reduces to the Wilson-Fisher family for $4-D\ll 1$. Unlike in previous work, our analytic continuation is non-perturbative. It is defined by using the conformal symmetry of fixed points. Recall that the free 4D scalar theory, critical 2D Ising model and, presumably, critical 3D Ising model possess such a symmetry, and we will assume that it survives for non-integer $D$ \footnote{This assumption can been checked order by order in the $\vareps$-expansion by exhibiting the conformal symmetry algebra of the anomalous dimension matrix of composite operators, as has been done at $O(\vareps)$ in \cite{Kehrein:1992fn,Kehrein:1994ff} and at $O(\vareps^2)$ in \cite{Kehrein:1995ia}.}. In integer dimensions, conformal symmetry leads to well-known constraints on correlation functions of local operators. For example, it fixes the correlator of four scalar operators up to a function $g(u,v)$ of conformal cross ratios. Since the number of independent cross ratios is the same (two) for any integer $D\ge 2$, it is natural to take the function $g(u,v)$ as the starting point for the analytic continuation. Recall that this function can be decomposed by using the operator product expansion (OPE) into a sum of conformal blocks corresponding to the exchanged operators. Inequivalent decomposition channels must produce the same four-point function, implying a crossing symmetry (`bootstrap') constraint. 
Furthermore, the conformal blocks are eigenfunctions of the quadratic Casimir operator, which depends on the space-time dimension $D$ analytically, so its eigenvalue equation can be solved treating $D$ as a free parameter. We take the decomposition of the function $g(u,v)$ into analytically-continued conformal blocks, together with the crossing symmetry constraint, as a definition of what it means to have conformal symmetry consistent with the OPE in non-integer dimensions.

\section{\label{sec:level3}Tracking Ising from 2D to 4D}

Following the logic of \cite{Rattazzi:2008pe}, we can place rigorous upper bounds on the dimension $\Delta_{\eps}$ of the first non-trivial scalar operator $\eps$ entering the OPE of the lowest dimension operator $\sigma$ with itself: 
\begin{equation}
\sigma \times \sigma \sim 1 + \eps +\cdots .
\end{equation}
This is done by formulating the bootstrap constraints on the four-point function $\langle\sigma\sigma\sigma\sigma\rangle$ as a linear program and solving it numerically by using the simplex method. For a given value of $\Delta_{\sigma}$, the linear program has no solution if $\Delta_\eps$ is sufficiently large. The details of our methodology will be elaborated on in a future publication \cite{El-Showk:2014dwa}. Notice that the only representations of $SO(D)$ that can occur in the conformal block decomposition of this four-point function are  symmetric traceless tensors of rank $\ell=0,2,4\ldots$. We analytically continue conformal blocks to non-integer $D$ separately for each $\ell$. We evaluate the blocks and their derivatives by using the expansion in radial coordinates \cite{Hogervorst:2013sma}, by the algorithm described in \cite{Hogervorst:2013kva}.

As first seen in \cite{Rychkov:2009ij} and \cite{ElShowk:2012ht}, the upper bound on $\Delta_{\eps}$ as a function of $\Delta_{\sigma}$ shows a sharp change of slope in both 2D and 3D. Within errors, the locations of these `kinks' agree with the known dimensions in the 2D and 3D Ising models. If we were to interpolate between these results by varying $D$, we would expect the position of the kink to evolve until it converges upon the free scalar theory in 4D, where no kink has been observed. One might then hypothesize that for some as yet unknown but likely fundamental reason the critical points of the Ising universality class lie exactly at the kink determined with the best possible accuracy.

\begin{figure*}[t]
\includegraphics[scale=1.38]{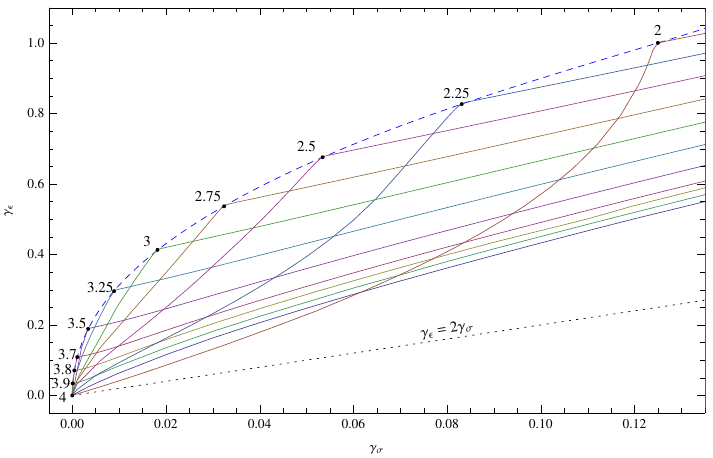}
\caption{\label{fig:summary} Upper bounds on $\gamma_\eps$ as a function of $\gamma_\sigma$, plotted for $D=2,2.25,\ldots,4$. For each $D<4$, the bound shows a kink, where a CFT belonging to the Ising model universality class is conjectured to live (black dots, fitted by the blue dashed curve). An example of theories in the bulk of the allowed region are Gaussian models, where $\gamma_\eps=2\gamma_\sigma$ (black dotted line).}
\end{figure*}

This intuition is indeed borne out by our analysis, summarized in Fig. \ref{fig:summary}, where we show upper bounds on $\Delta_{\eps}$ for different values of $D$. We plot the results in terms of anomalous dimensions, defined as the difference between an operator's scaling dimension and the 
scaling dimension of the corresponding field in the free scalar theory in $D$ dimensions:
\begin{eqnarray}
&& \gamma_\sigma \equiv \Delta_\sigma - \Delta_\phi  = \Delta_\sigma - (D-2)/2\nonumber ,\\
&& \gamma_\eps \equiv \Delta_\eps - \Delta_{\phi^2}  = \Delta_\eps - (D-2) .
\end{eqnarray}
As expected, all the bounds possess kinks, which become sharper as $D \rightarrow 4$. These kinks are clearly special points in the space of scaling dimensions. By construction, for the $\langle \sigma\sigma\sigma\sigma\rangle$ correlator crossing symmetry has a solution anywhere below the bound. We conjecture that at the kinks this solution can be extended to all the correlators of the theory; i.e., there is a full-fledged CFT corresponding to these operator dimensions.

\begin{figure*}
\includegraphics[scale=1.38]{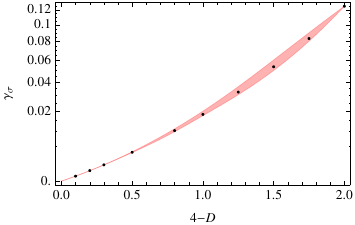}
\includegraphics[scale=1.38]{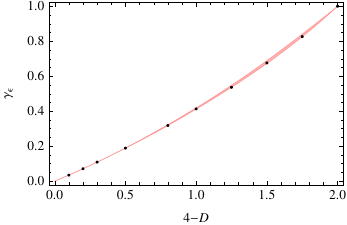}
\caption{\label{fig:eps exp} Black dots: The anomalous dimensions corresponding to the kinks in Fig.~\ref{fig:summary}. Red bands: The same dimensions determined by Borel-resumming the $\vareps$-expansion series \cite{LeGuillou:1987ph}. Since $\gamma_\sigma=O(\vareps^2)$, we use a square root scale on the $\gamma_\sigma$ axis.}
\end{figure*}

To test this conjecture, we compare the positions of the kinks with the $\vareps$-expansion. We use the results of Ref.~\cite{LeGuillou:1987ph}, where the $\vareps$-expansion was Borel-resummed for a number of dimensions between $2$ and $4$, imposing agreement with the exactly-known 2D critical exponents as a boundary condition. As Fig.~\ref{fig:eps exp} shows, we find excellent agreement within the stated error bars \footnote{Although the $O(\vareps^5)$ terms used in \cite{LeGuillou:1987ph} contain a mistake since then corrected in \cite{Kleinert:1991rg}, the induced error is negligible \cite{Guida:1998bx}.}. For $\vareps\lesssim 0.5$ errors due to ambiguities in the resummation procedure are negligible, and our points precisely track the $\vareps$-expansion curve \footnote{The lowest order $\vareps$-expansion result for $\gamma_\eps$ was reproduced by the conformal bootstrap methods long ago in \cite{Polyakov:1974gs}.}. 

\section{\label{sec:level4} Central Charges }

A future goal of this study is to be able to track the spectrum of the CFTs at the kinks from the free 4D scalar to the Ising model in 2D. As a preliminary step in this program we investigate the value of the central charge, defined as the coefficient in the two-point function of the canonically-normalized stress tensor. We normalize the two-point function so that the free scalar central charge is $c_T^{\text{free}}=D/(D-1)$ \cite{Osborn:1993cr}.

Hence, for any dimension $D$, we assume the maximal gap allowed by our dimension bound and we extract the solution to the crossing symmetry constraint for several values of $\gamma_\sigma$ around the kink. For $\gamma_\sigma$ fixed and $\gamma_\eps$ approaching the bound from below, we observe that the dimensions and OPE coefficients of low-lying operators in the solution approach finite limits. Such a behavior was previously speculated in Refs.~\cite{Poland:2010wg,ElShowk:2012ht}; a dual version of the same phenomenon was demonstrated in the 2D case in Ref.~\cite{ElShowk:2012hu}. Here we focus on the stress tensor of the theory, identified as the symmetric traceless rank-two tensor operator of dimension $D$. By inspection, an operator with such quantum numbers turns out to be always present in the limiting solution.
The value $\lambda_{D,2}^2$ of its OPE coefficient squared is then related to the central charge by an inverse relation:
\begin{equation}
c_T =  ({\Delta_\sigma^2}/{\lambda_{D,2}^2}) c_T^{\text{free}} \,,
\label{ref:c_T}
\end{equation}
where $\lambda_{D,2}^2$ is extracted by normalizing the conformal blocks as in Ref.~\cite{DO3,ElShowk:2012ht}.

\begin{figure}
\includegraphics[scale=1.38]{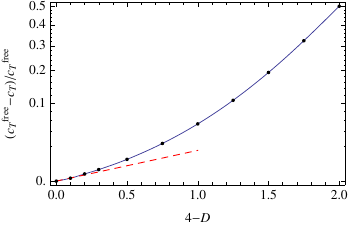} 
\caption{\label{fig:c_T} Black dots: The normalized central charge difference (on a square root scale) corresponding to the kinks in Fig.~\ref{fig:summary}, interpolated by the blue curve. The dashed red line is the lowest-order $\vareps$-expansion prediction.}
\end{figure}

For each $D$, Eq.~\reef{ref:c_T} gives $c_T$ as a function of $\gamma_\sigma$, the dependence coming both from the $\Delta_\sigma$ in Eq.~(\ref{ref:c_T}) and from the fact that $\lambda_{D,2}^2$ is determined from the limiting solution which is a nontrivial function of $\gamma_\sigma$. For any $D$, the dependence of $c_T$ on $\gamma_\sigma$ is qualitatively similar to Fig.~11 in Ref.~\cite{ElShowk:2012ht}. Namely, it turns out that $c_T$ has a minimum for $\gamma_\sigma$ at the kink, which we would like to identify as the value of $c_T$ for the CFT living at the kink \footnote{In fact, exactly the same values of the central charge for boundary solutions can be obtained by computing bounds on OPE coefficients as in \cite{ElShowk:2012ht}.}. For $D=2$ this agrees very precisely with the exact 2D Ising model value \cite{ElShowk:2012hu,Vichi-thesis}. Interestingly, $c_T<c_T^{\text{free}}$ for all $2\le D< 4$, although Zamolodchikov's $c$-theorem mandates this only for $D=2$. In Fig.~\ref{fig:c_T} we plot the normalized difference $(c_T^{\text{free}}-c_T)/c_T^{\text{free}}$ as a function of $\vareps$. This represents our prediction for the central charge as the space-time dimension changes from 4 to 2. The dashed line in the same plot shows the $\vareps$-expansion prediction \cite{Hathrell:1981zb,Jack:1983sk,Cappelli:1990yc,Petkou:1994ad}:
\beq
(c_T^{\text{free}}-c_T)/c_T^{\text{free}}=5\vareps^2/{324}+\cdots.
\eeq
The agreement is good for $\vareps\lesssim 0.3$, but for larger values the unknown higher order corrections must be significant.

\section{\label{sec:level5} Discussion }
The results of this study show clearly that the family of CFTs conjectured by Wilson and Fisher can be identified at the non-perturbative level using the conformal bootstrap. The bootstrap predictions match well with the best estimates from the Borel-resummed $\vareps$-expansion for $\gamma_{\sigma}$ and $\gamma_{\eps}$, and greatly surpass previous perturbative computations of the central charge. We are optimistic that we will soon obtain precise bootstrap predictions in this family of CFTs for the full low-lying spectrum of operator dimensions and OPE coefficients. 

There are still a number of important questions that remain to be answered -- why does this family of CFTs occupy a special place in the space allowed by crossing symmetry and unitarity? Can we gain a better analytic understanding of the transitions across the kinks? Can one apply similar techniques on other correlators to learn about the $\mathbb{Z}_2$-odd spectrum? Finally, can one adapt similar techniques to identify and learn about 
CFTs living in the interior? We hope that these and related questions can be addressed in future work.

While in this paper we only studied $2\le D \le4$, it should be straightforward and interesting to extend our analysis to $1<D<2$, where the line of Wilson-Fisher fixed points is believed to continue \cite{LeGuillou:1987ph}, and to connect to the conformal bootstrap studies in $D=1$ \cite{Gaiotto:2013nva}. From the CFT point of view, the $D\to1$ limit na\"ively looks discontinuous, since in $D=1$ we have only one cross-ratio and no spin. This issue deserves a detailed study.

\section*{Acknowledgements}
S.R. is grateful to J.~Cardy for useful discussions, and to H.W. Diehl for informing him about Refs.~\cite{Kehrein:1992fn,Kehrein:1994ff,Kehrein:1995ia}.
We thank the organizers and participants for the stimulating environment at the ``Back to the Bootstrap 3'' conference at CERN. The work of S.E. was supported by the French ANR Contract No.
05-BLAN-NT09-573739, the ERC Advanced Grant No. 226371 and the ITN
program PITN-GA-2009-237920.
M.P. is supported by DOE Grant  No. DE-SC0010010-Task A.
D.P. thanks the Galileo Galilei Institute for Theoretical Physics and the INFN for hospitality and partial support
during the completion of this work. The work of S.R. is supported in part by the European Program ``Unification in the LHC Era'', 
contract PITN-GA-2009-237920 (UNILHC). The work of D.S.D. is supported by DOE grant number DE-SC0009988.  D.S.D. thanks SLAC for hospitality while this work was being completed.  This research used resources of the National Energy Research
Scientific Computing Center, which is supported, as the work of A.V., by the Office of
Science of the U.S. Department of Energy under Contract No.
DE-AC02-05CH1123.

\bibliography{Biblio}{}
\bibliographystyle{utphys}

\end{document}